\documentclass[twocolumn, showpacs,prl,floatfix]{revtex4}%
\usepackage{graphicx}
\usepackage{dcolumn}
\usepackage{bm}
\usepackage{amsmath}
\usepackage{amsfonts}
\usepackage{amssymb}%
\providecommand{\U}[1]{\protect\rule{.1in}{.1in}}
\begin{document}
\author{R. Cabrera, W. E. Baylis\thanks{Corresponding author. \emph{E-mail address}:
baylis@uwindsor.ca}}
\affiliation{Department of Physics, University of Windsor, Windsor, ON N9B 3P4. Canada}
\title{Average Fidelity in $n$-Qubit systems}

\begin{abstract}
This letter generalizes the expression for the average fidelity of single
qubits, as found by Bowdrey et al. \cite{Bowdrey2002}, to the case of $n$
qubits. We use a simple algebraic approach with basis elements for the
density-matrix expansion expressed as Kronecker products of $n$ Pauli spin
matrices. An explicit integration over initial states is avoided by invoking
the invariance of the state average under unitary transformations of the
initial density matrix. The results have applications to measurements of
quantum information, for example in ion-trap and NMR experiments.

\end{abstract}

\pacs{03.67.-a; 82.56.-b; 03.65.Fd}
\maketitle

The fidelity, usually defined\cite{Jozsa94} as
\begin{equation}
F(\rho,\rho^{\prime})=\left(  \operatorname*{tr}\sqrt{\rho^{1/2}\rho^{\prime
}\rho^{1/2}}\right)  ^{2} \label{Uhlmann}%
\end{equation}
(the square of the Uhlmann \cite{Uhlmann1976} formula), measures the
correlation between two states represented by density matrices $\rho$ and
$\rho^{\prime}.$ The evaluation of fidelity is important in the storage,
processing and communication of quantum information\cite{Grangier05,Deuar00},
for example in optical qubits\cite{OBrien05}, ion-trap
measurements\cite{NIST2006} or NMR experiments, especially those with
high-purity state preparation\cite{Anwar04}. Expression (\ref{Uhlmann}) is
simplified if one of the density matrices, say $\rho,$ is in a pure state:
$\rho=\left\vert \Psi\right\rangle \left\langle \Psi\right\vert $ . Then
$\rho$ is idempotent so that $\rho^{1/2}=\rho$ and
\begin{align}
F(\rho,\rho^{\prime})  &  =\left(  \operatorname*{tr}\sqrt{\left\vert
\Psi\right\rangle \left\langle \Psi\right\vert \rho^{\prime}\left\vert
\Psi\right\rangle \left\langle \Psi\right\vert }\right)  ^{2}=\left\langle
\Psi\right\vert \rho^{\prime}\left\vert \Psi\right\rangle \nonumber\\
&  =\operatorname*{tr}(\rho\rho^{\prime}). \label{fidel1}%
\end{align}
This is a convenient form for calculating the fidelity of two output states
coming from the same initial pure state $\rho_{0}=\left\vert \Psi
_{0}\right\rangle \left\langle \Psi_{0}\right\vert $, where $\rho$ is the
result of a unitary transformation that preserves the purity of the state, and
$\rho^{\prime}$ comes from a more general trace-preserving linear map
$\mathcal{M}$:
\begin{align}
\rho &  =U\rho_{0}U^{\dagger}\nonumber\\
\rho^{\prime}  &  ={\mathcal{M}}(\rho_{0}). \label{output}%
\end{align}
The average fidelity is defined by averaging over all possible initial states
$\rho_{0}:$
\begin{equation}
\langle F\rangle=\langle F(\rho,\rho^{\prime})\rangle_{\rho_{0}}%
=\langle\operatorname*{tr}(\rho\rho^{\prime})\rangle_{\rho_{0}}~.
\end{equation}

In the case of a single qubit, the average can be found by integrating over
the Bloch sphere \cite{Bowdrey2002}
\begin{equation}
\langle F\rangle=\frac{1}{4\pi}\int\operatorname*{tr}(\rho\rho^{\prime
})d\Omega,
\end{equation}
where $\rho_{0}$ is expanded in Pauli spin matrices $\sigma_{j}$ as
\begin{equation}
\rho_{0}=\frac{1}{2}(\mathbf{1}+\sigma_{1}\sin\theta\cos\phi+\sigma_{2}%
\sin\theta\sin\phi+\sigma_{3}\cos\theta)
\end{equation}
and $d\Omega=\sin\theta d\theta d\phi$ is an element of solid angle. Bowdrey
\emph{et al.} \cite{Bowdrey2002} showed how the integration can be exactly
replaced by a finite sum in the case of a single qubit. For the more general
$n$-qubit case, the corresponding integration is difficult to envision.
Nevertheless, a generalization to \textquotedblleft qudit\textquotedblright%
\ systems of $d=N$ independent states was developed by Bagan \emph{et al.}
\cite{Bagan2003}, by expanding ~$\rho_{0}$ in the generators of $SU\left(
N\right)  $ and integrating the fidelity over an invariant Haar measure for
unitary operators. In a different approach exploiting global symmetries to
relate the integral over pure states to invariant traces, Pedersen \emph{et
al.} \cite{Pedersen2007} were able to express the fidelity for any completely
positive trace-preserving linear map with a known Kraus-operator form
\cite{Nielsen2000}, giving a formula equivalent to one that Emerson \emph{et
al.} \cite{Emerson2005} found. In this letter, we present a simple approach
with an explicit polarization basis that avoids the integration over the space
of pure states, and with this approach we extend the Bowdrey \emph{et al.}
result to the case of $n$-qubit systems and show how a basis of pure states
can be constructed for use in practical applications.

To explain our approach, we first rederive the result for a single-qubit
systems. The density matrix $\rho$ of a single qubit can be written
\cite{Bay03a} in terms of the polarization vector $\mathbf{P,}$ represented
with components $P^{j}$ on the Pauli matrices
\begin{equation}
\mathbf{P}=\sum_{j}P^{j}\sigma_{j}\equiv P^{1}\sigma_{1}+P^{2}\sigma_{2}%
+P^{3}\sigma_{3},
\end{equation}
as
\begin{equation}
\rho=\frac{1}{2}\left(  \mathbf{1}+\mathbf{P}\right)  . \label{rho0}%
\end{equation}
When $\rho$ is in a pure state, $\mathbf{P}$ lies on the surface of the unit
sphere
\begin{equation}
\mathbf{P}^{2}=(P^{1})^{2}+(P^{2})^{2}+(P^{3})^{2}=1,
\end{equation}
whereas for a general mixed state, it lies within the sphere: $\mathbf{P}%
^{2}<1.$ When $\mathbf{P}=0,$ the state is unpolarized (contains no
polarization information).

If we use expression (\ref{rho0}) for the initial state $\rho_{0},$ the output
states (\ref{output}) are
\begin{align}
\rho &  =\frac{1}{2}(\mathbf{1}+U\mathbf{P}U^{\dagger})\\
\rho^{\prime}  &  =\frac{1}{2}({\mathcal{M}}(\mathbf{1})+\mathcal{M}%
(\mathbf{P})),
\end{align}
and the fidelity becomes
\begin{align}
F  &  =\frac{1}{4}\operatorname*{tr}\left[  (\mathbf{1}+U\mathbf{P}U^{\dagger
})({\mathcal{M}(\mathbf{1})}+\mathcal{M}(\mathbf{P})\right] \label{Fidel1q}\\
&  =\frac{1}{4}\operatorname*{tr}\left[  \mathbf{1}+\sum_{j}P^{j}U\sigma
_{j}U^{\dagger}\left(  {\mathcal{M}}(\mathbf{1})+\sum_{k}P^{k}{\mathcal{M}%
}(\sigma_{k})\right)  \right]  .\nonumber
\end{align}
We want to average $F$ over initial states, that is over all possible
directions of $\mathbf{P.}$ Since the coefficients $P^{j}$ are independent and
as likely to be positive as negative, the average values of $P^{j}$ and of
$P^{j}P^{k},\ j\neq k,$ vanish. Furthermore, since $\rho_{0}$ is assumed to be
in a pure state,
\begin{equation}
\langle(P^{1})^{2}+(P^{2})^{2}+(P^{3})^{2}\rangle=1,
\end{equation}
and by symmetry the contribution of each component must be the same,
\begin{equation}
\langle(P^{1})^{2}\rangle=\langle(P^{2})^{2}\rangle=\langle(P^{3})^{2}%
\rangle=\frac{1}{3}. \label{symm1}%
\end{equation}
It follows that%
\begin{equation}
\langle P^{j}P^{k}\rangle=\frac{1}{3}\delta^{jk}%
\end{equation}
The average fidelity (\ref{Fidel1q}) is thus
\begin{align}
\langle F\rangle &  =\frac{1}{4}\operatorname*{tr}\left(  \mathbf{1}+\frac
{1}{3}\sum_{j}U\sigma_{j}U^{\dagger}{\mathcal{M}}(\sigma_{j})\right)
\nonumber\\
&  =\frac{1}{3}\sum_{j}\operatorname*{tr}\left(  \frac{\mathbf{1}}{4}%
+U\frac{\sigma_{j}}{2}U^{\dagger}{\mathcal{M}}(\frac{\sigma_{j}}{2})\right)  .
\label{Fav1q}%
\end{align}
The Pauli matrices can be written in terms of the pure states $\rho_{\pm
j}\equiv\frac{1}{2}\left(  \mathbf{1}\pm\sigma_{j}\right)  $ as
\begin{align}
\frac{\sigma_{j}}{2}  &  =\frac{\mathbf{1}+\sigma_{j}}{2}-\frac{\mathbf{1}}%
{2}=\rho_{j}-\hat{\rho}\\
&  =\frac{\mathbf{1}}{2}-\frac{\mathbf{1}-\sigma_{j}}{2}=\hat{\rho}-\rho_{-j},
\end{align}
where $\hat{\rho}=\mathbf{1}/2$ is the unpolarized (maximally mixed) state$.$
Then, $\left\langle F\right\rangle $ (\ref{Fav1q}) becomes%
\begin{align}
\langle F\rangle &  =\frac{1}{3}\sum_{j}\operatorname*{tr}\left(
\frac{\mathbf{1}}{2}+U\rho_{j}U^{\dagger}{\mathcal{M}}(\rho_{j})-U\rho
_{j}U^{\dagger}\mathcal{M}(\hat{\rho})\right) \\
&  =\frac{1}{3}\sum_{j}\operatorname*{tr}\left(  \frac{\mathbf{1}}{2}%
+U\rho_{-j}U^{\dagger}{\mathcal{M}}(\rho_{-j})-U\rho_{-j}U^{\dagger
}\mathcal{M}(\hat{\rho})\right)  .\nonumber
\end{align}
The average of these expressions gives the form found by Bowdrey \emph{et al.}
\cite{Bowdrey2002}%
\begin{align}
\langle F\rangle &  =\frac{1}{6}\sum_{j}Tr(\mathbf{1}-U(\rho_{j}+\rho
_{-j})U^{\dagger}\mathcal{M}(\hat{\rho})+\nonumber\\
&  U\rho_{j}U^{\dagger}{\mathcal{M}}(\rho_{j})+U\rho_{-j}U^{\dagger
}{\mathcal{M}}(\rho_{-j}))\\
&  =\frac{1}{6}\sum_{j}Tr(U\rho_{j}U^{\dagger}{\mathcal{M}}(\rho_{j}%
)+U\rho_{-j}U^{\dagger}{\mathcal{M}}(\rho_{-j})).\nonumber
\end{align}

This result is extended to $n$-qubits by defining an appropriate basis for the
density matrix expansion. First however, we look again at the symmetry
argument. While hardly anyone would question the equality of the average
square components of the polarization vector in Eq. (\ref{symm1}), we want to
invoke a rigorous proof that does not depend explicitly on the integration
over all states, and that we can extend to systems of two or more qubits. The
proof relies on the recognition that an average over all initial states is
invariant under unitary transformation of $\rho_{0}$. Such transformations are
a one-to-one mappings of each state of the system onto another (or the same)
state. For a single qubit, unitary transformations include rotations on the
Bloch sphere \cite{Bay03a} that could, for example, rotate $\mathbf{P}%
=P^{1}\sigma_{1}+P^{2}\sigma_{2}+P^{3}\sigma_{3}$ into
\begin{equation}
\mathbf{P}\rightarrow R\mathbf{P}R^{\dag}=\mathbf{P}^{\prime}=P^{1}\sigma
_{2}-P^{2}\sigma_{1}+P^{3}\sigma_{3},
\end{equation}
where $R$ is the unitary \textquotedblleft rotor\textquotedblright%
\ $R=\exp\left(  \sigma_{2}\sigma_{1}\pi/4\right)  =\exp\left(  -i\sigma
_{3}\pi/4\right)  .$ Since the average of any expression over initial states
is invariant under such a transformation, we must have $\left\langle
P^{1}P^{2}\right\rangle =\left\langle -P^{2}P^{1}\right\rangle $ and
$\left\langle \left(  P^{1}\right)  ^{2}\right\rangle =\left\langle \left(
P^{2}\right)  ^{2}\right\rangle $ as argued above. As we see below, there is a
richer selection of possible unitary transformations in systems of more qubits.

For two qubits the density matrix can be expanded in a polarization basis
$\left\{  \mathbf{f}_{j}\right\}  $ as
\begin{equation}
\rho=\frac{1}{4}(1+\sum_{j=1}^{15}w^{j}\mathbf{f}_{j}),
\end{equation}
with basis elements defined in terms of a Kronecker product of Pauli matrices
\begin{equation}
\mathbf{f}_{j}=\sigma_{\mu}\otimes\sigma_{\nu},\,\,\,\,\mu,\nu=0,1,2,3,
\label{polbasis}%
\end{equation}
where $\sigma_{0}$ is the $2\times2$ identity matrix and $\sigma_{0}%
\otimes\sigma_{0}$ is excluded from the sum. The $\mathbf{f}_{j}$ are
traceless and satisfy the orthonormality condition
\begin{equation}
\operatorname*{tr}(\mathbf{f}_{i}\mathbf{f}_{j})=4\delta_{ij}~. \label{orthon}%
\end{equation}
Consequently, for any states $\rho_{1},\rho_{2},$
\begin{align}
\operatorname*{tr}\left(  \rho_{1}\rho_{2}\right)   &  =\frac{1}%
{16}\operatorname*{tr}\left[  \left(  \mathbf{1}+\sum_{j=1}^{15}w_{1}%
^{j}\mathbf{f}_{j}\right)  \left(  \mathbf{1}+\sum_{k=1}^{15}w_{2}%
^{k}\mathbf{f}_{k}\right)  \right] \nonumber\\
&  =\frac{1}{4}\left(  1+\sum_{j=1}^{15}w_{1}^{j}w_{2}^{j}\right)  .
\end{align}
If $\rho_{1},\rho_{2}$ are the output states $\rho,\rho^{\prime}$
(\ref{output}), the fidelity (\ref{fidel1}) is%
\begin{align}
F  &  =\operatorname*{tr}(U\rho_{0}U^{\dagger}{\mathcal{M}}(\rho
_{0}))\label{F2q}\\
&  =\frac{1}{16}\operatorname*{tr}\!\left[  \!\left(  \mathbf{1}+\sum
_{j=1}^{15}w_{0}^{j}U\mathbf{f}_{j}U^{\dag}\right)  \!\left(  {\mathcal{M}%
}\left(  \mathbf{1}\right)  +\sum_{k=1}^{15}w_{0}^{k}{\mathcal{M}}\left(
\mathbf{f}_{k}\right)  \right)  \!\right] \nonumber
\end{align}
For any pure state such as $\rho_{0},$%
\begin{equation}
\operatorname*{tr}\left(  \rho_{0}^{2}\right)  =\operatorname*{tr}\left(
\rho_{0}\right)  =1=\frac{1}{4}\left(  1+\sum_{j=1}^{15}\left(  w_{0}%
^{j}\right)  ^{2}\right)
\end{equation}
and thus
\begin{equation}
\sum_{j=1}^{15}\left(  w_{0}^{j}\right)  ^{2}=3. \label{w-cond}%
\end{equation}
More generally, for a state that may be mixed,
\begin{equation}
\sum_{j=1}^{15}\left(  w^{j}\right)  ^{2}\leq3
\end{equation}
with the equality holding only when the state is pure. The average of
(\ref{w-cond}) can be reduced by the symmetry that such expressions are
invariant under any unitary transformation of $\rho_{0},$ as discussed in the
single-qubit case, above. There are unitary transformations that rotate any
pair of polarization basis elements $\mathbf{f}_{j}$ and $\pm\mathbf{f}_{k}$
(\ref{polbasis}) into each other. For example,%
\begin{align*}
\sigma_{1}\otimes1  &  \rightarrow\left(  R\otimes1\right)  \left(  \sigma
_{1}\otimes1\right)  \left(  R^{\dag}\otimes1\right)  =\sigma_{2}\otimes1\\
\sigma_{2}\otimes1  &  \rightarrow\left(  R\otimes1\right)  \left(  \sigma
_{2}\otimes1\right)  \left(  R^{\dag}\otimes1\right)  =-\sigma_{1}\otimes1
\end{align*}
and%
\[
1\otimes\sigma_{1}\rightarrow\left(  1\otimes R\right)  \left(  1\otimes
\sigma_{1}\right)  \left(  1\otimes R^{\dag}\right)  =1\otimes\sigma_{2},
\]
with $R=\exp\left(  \sigma_{2}\sigma_{1}\pi/4\right)  =\exp\left(
-i\sigma_{3}\pi/4\right)  ,$ are single-qubit rotations, whereas%
\[
\sigma_{1}\otimes1\rightarrow R_{12}\sigma_{1}\otimes1R_{12}^{\dag}=\sigma
_{2}\otimes\sigma_{1},
\]
with $R_{12}=\exp\left(  \sigma_{2}\sigma_{1}\otimes\sigma_{1}\pi/4\right)  ,$
is a coupled-qubit rotation. Such coupled-qubit rotations are required to
transform an unentangled state to an entangled one and \emph{vice versa.} Of
course single- and coupled-qubit rotations can be sequenced to yield still
other unitary transformations. As a result of the invariance of state averages
under all unitary transformations, the average of $\left(  w_{0}^{j}\right)
^{2}$ must be the same for every $j,$ and we can reduce the sum for a pure
state $\rho_{0}$ by%
\begin{equation}
3=\langle\sum_{j=1}^{15}\left(  w_{0}^{j}\right)  ^{2}\rangle=15\langle
(w_{0}^{j})^{2}\rangle.
\end{equation}
More generally
\begin{equation}
\langle w_{0}^{i}w_{0}^{j}\rangle=\delta^{ij}\langle(w_{o}^{j})^{2}%
\rangle=\frac{1}{5}\delta^{ij}.
\end{equation}
Since there are also unitary transformations that simply change the sign of
the polarization components, the average $\rho_{0}$ is unpolarized:
$\left\langle w_{0}^{j}\right\rangle =0.$ Combining these relations, the
average of the fidelity (\ref{F2q}) reduces to%
\begin{equation}
\left\langle F\right\rangle =\frac{1}{16}\left[  4+\frac{1}{5}\sum_{j=1}%
^{15}\operatorname*{tr}\left[  U\mathbf{f}_{j}U^{\dag}{\mathcal{M}}\left(
\mathbf{f}_{j}\right)  \right]  \right]  \label{Fav2q}%
\end{equation}

For the general $n$-qubit case, the basis element $\mathbf{f}_{j}$ is a
Kronecker product of $n$ Pauli spin matrices $\sigma_{\mu}$. The polarization
basis (excluding the identity element) contains $N^{2}-1$ elements
$\mathbf{f}_{j}$ with $N=2^{n},$ and the density matrix can be expanded as
\begin{equation}
\rho=\frac{1}{N}(1+\sum_{j=1}^{N^{2}-1}w^{j}\mathbf{f}_{j}).
\end{equation}
The trace of the square of a pure-state density matrix is $\operatorname*{tr}%
(\rho_{0}^{2})=1,$ and the $\mathbf{f}_{j}$ satisfy%
\begin{equation}
\operatorname*{tr}\left(  \mathbf{f}_{j}\mathbf{f}_{k}\right)  =N\delta_{jk}~.
\end{equation}
Consequently,
\begin{align*}
1  &  =\frac{1}{N^{2}}\operatorname*{tr}\left[  \left(  1+\sum_{j=1}^{N^{2}%
-1}w_{0}^{j}\mathbf{f}_{j}\right)  \left(  1+\sum_{k=1}^{N^{2}-1}w_{0}%
^{k}\mathbf{f}_{k}\right)  \right] \\
&  =\frac{1}{N}+\frac{1}{N}\sum_{jk}w_{0}^{j}w_{0}^{k}\delta_{jk}=\frac{1}%
{N}\left(  1+\sum_{j=1}^{N^{2}-1}\left(  w_{0}^{j}\right)  ^{2}\right)  ,
\end{align*}
which implies%
\begin{equation}
\sum_{j=1}^{N^{2}-1}\left(  w_{0}^{j}\right)  ^{2}=N-1~. \label{sumw2}%
\end{equation}
Since, as in the 2-qubit case, there are unitary transformations that
interchange the polarization elements $\mathbf{f}_{j}$ or change their signs,
the average of (\ref{sumw2}) gives
\[
\langle(w_{0}^{j})^{2}\rangle=\frac{N-1}{N^{2}-1}=\frac{1}{1+N},
\]
and the average including the cross terms is%
\begin{equation}
\langle w_{0}^{i}w_{0}^{j}\rangle=\frac{1}{1+N}\delta^{ij}.
\end{equation}
The average of the fidelity $F=\operatorname*{tr}(U\rho_{0}U^{\dagger
}{\mathcal{M}}(\rho_{0}))$ is thus generalized from (\ref{Fav2q}) to%
\begin{align}
\langle F\rangle &  =\frac{1}{N^{2}}\operatorname*{tr}\left[  \mathbf{1}%
+\sum_{jk}\langle w_{0}^{j}w_{0}^{k}\rangle U\mathbf{f}_{j}U^{\dagger
}{\mathcal{M}}(\mathbf{f}_{k})\right] \nonumber\\
&  =\frac{1}{N}+\frac{1}{(N+1)N^{2}}\sum_{j}^{N^{2}-1}\operatorname*{tr}%
\left[  U\mathbf{f}_{j}U^{\dagger}{\mathcal{M}}(\mathbf{f}_{j})\right]  .
\label{Favnq}%
\end{align}
This result confirms the qudit-fidelity relation derived independently by
Bagan \emph{et al.} \cite{Bagan2003}

The remaining step is to express the basis elements $\mathbf{f}_{j}$ in terms
of a set of pure-state density operators. Such operators are primitive
idempotents \cite{Bay03a,Lounesto2001} in the algebra, and in the $n$-qubit
case they can be expressed as products of $n$ commuting simple idempotents. In
the 2-qubit case, we can use the simple idempotents%
\begin{equation}
P_{\pm\mu\nu}=\frac{1}{2}\left(  \mathbf{1}\pm\sigma_{\mu}\otimes\sigma_{\nu
}\right)
\end{equation}
and note the following identities to express $\mathbf{f}_{j}=\sigma_{\mu
}\otimes\sigma_{\nu}$ as a linear combination of pure-state density matrices
\begin{align}
\sigma_{\mu}\otimes\sigma_{\nu}  &  =\left(  P_{+\mu\nu}-P_{-\mu\nu}\right)
\left(  P_{+\nu\mu}+P_{-\nu\mu}\right)  ,~\mu>\nu\\
&  =\left(  P_{+\mu0}-P_{-\mu0}\right)  \left(  P_{+0\nu}-P_{-0\nu}\right)
,~\mu>0,~\nu>0~.\nonumber
\end{align}
A similar procedure is possible for the general $n$-qubit case. Although the
explicit expressions become rather lengthy, the procedure is straightforward.

\subsection*{Acknowledgements}

The authors thank the Natural Sciences and Engineering Research Council of
Canada for support of this work.

\end{document}